\documentstyle[11pt,IAUS212,twoside,graphicx,longtable]{article}

\def\edcomment#1{\iffalse\marginpar{\raggedright\sl#1\/}\else\relax\fi}
\reversemarginpar

\begin{document}
\vspace*{1cm}
\title{Aspherical Supernovae Explosions}
 \author{P. H\"oflich $^1$,                A. Khokhlov $^2$, L. Wang $^3$, J.C. Wheeler $^1$, D. Baade$^4$  }
\affil{$^1$ Dept, of Astronomy, University of Texas, Austin, TX 78681, USA}
\affil {$^2$ Naval Research Lab, Washington DC, USA}
\affil {$^3$ Lawrence Berkeley Lab, 1 Cyclotron Rd, Berkeley, CA 94720, USA}
\affil{$^4$ ESO,  Karl-Schwarzschild-Str. 2, D-85748 Garching, Germany}

\begin{abstract}
 Core collapse supernovae (SN) are the final stages of stellar evolution in 
massive stars during which the central region collapses, forms a neutron star 
(NS), and the outer layers are ejected. Recent explosion scenarios assumed that 
the ejection is due to energy deposition by neutrinos into the envelope but 
detailed models do not produce powerful explosions. There is new and mounting evidence
for an asphericity and, in particular, for axial symmetry in several supernovae which
may be hard to reconcile within the spherical picture.
  This evidence includes the observed high polarization and its variation with time,
pulsar kicks, high velocity iron-group and intermediate-mass elements material 
observed in remnants,  direct observations of the debris of SN1987A etc.
 Some of the new evidence is discussed in more detail.
 To be in agreement with the observations, any successful mechanism must invoke
some sort of axial symmetry for the explosion.
We consider jet-induced/dominated explosions of
core collapse supernovae.
Our study is based on detailed 3-D hydrodynamical
and radiation transport models. We find that
the observations can be explained by low velocity, massive jets which
stall well within the SN envelope. Such outflows may be produced by  MHD-mechanisms,
convective dominated accretion disks on the central object or asymmetric neutrino emissions.
Asymmetric density/chemical distributions and, for SN2002ap,
off-center energy depositions have been identified as crucial for the interpretation of 
the polarization.
\end{abstract}

\section{Introduction}
\vskip -0.2cm
There is a general agreement that the explosion of a massive star is caused
by the collapse of its central parts into a neutron star or, for massive 
progenitors, into a black hole.
The mechanism of the energy deposition into the envelope
is still debated.  The process likely involves the bounce and 
the formation of the prompt shock (e.g. Van Riper 1978),
 radiation of the energy in the form of
neutrinos (e.g. Bowers \& Wilson 1982), and the interaction of the neutrino with the material of
the envelope and  various types of convective motions ( e.g. Herant et al. 1994, Burrows et al. 1995,
 M\"uller \& Janka 1997, Fryer \& Waren 2002), rotation (e.g. LeBlanc \& Wilson 1970,
M\"onchmeyer et al. 1991)
and magnetic fields (e.g. LeBlanc \& Wilson 1970, Bisnovati-Kogan 1971, Symbalisty 1984).
 Currently, the most favored mechanism invokes the neutrino deposition but it is neither clear whether
it can provide powerful explosions, can account for large scale asymmetries or for large kick velocities as
observed in neutron stars (see below).
Likely, a quantitative model of the core collapse must eventually include all the elements mentioned above
(H\"oflich et al. 2001).
 In this paper we study the effects and observational consequences of an asymmetric, jet-like
deposition of energy inside the envelope of SN.                                

\section{Evidence for Asymmetry}
\vskip -0.2cm
In recent years, there has been a mounting evidence that the explosions of massive stars (core
collapse supernovae) are highly aspherical. 
(1) The spectra of core-collapse supernovae (e.g., SN87A, SN93J, SN94I, SN99em, SN02ap)
 are significantly polarized at a level of 0.5 to 3 \% 
  (e.g. Fig. 1, M\'endez et al. 1988; Cropper et al. 1988; H\"oflich 1991; Jeffrey 1991;
 Wang et al. 1996) indicating aspherical envelopes by factors of up to 2 (Fig.2).
The degree of polarization  tends to vary inversely with the mass of the hydrogen
envelope, being maximum for Type Ib/c events with no hydrogen 
(Wang et al. 2001). For SNeII, Leonard et al. (2000) and Wang et al. (2001) showed  that the
polarization and, thus, the asphericity increase  with time. Both trends suggest a connection
of the asymmetries with the central engine.
 For supernovae with a good time and wavelength
coverage, the orientation of the polarization vector tends to stay constant
both in time and with wavelength.  This implies that there is a  global symmetry
axis in the ejecta (Leonard et al. 2001, Wang et al. 2001).
(2) Observations of SN~1987A showed that radioactive material was brought
to the hydrogen rich layers of the ejecta very quickly during the explosion
(e.g Lucy 1988).
(3) The remnant of the Cas~A supernova shows rapidly moving oxygen-rich
matter outside the nominal boundary of the remnant             
                 and  evidence for two oppositely directed jets of
high-velocity material (Fesen \& Gunderson 1997). 
(4) Recent X-ray observations with the CHANDRA satellite have shown an unusual
distribution of iron and silicon group elements with large
scale asymmetries in Cas~A (Huges et al. 2000).
(5) After the explosion, neutron stars are observed with high
velocities, up to 1000  km/s (Strom et al. 1995).
(6) Direct HST-images from June 11,2000, are able to resolve the inner debris of SN1987A showing
its prolate geometry with an axis ratio of $\approx 2$, and spectra indicate
that the  products of stellar burning (O, Ca, etc.) are concentrated in the equatorial plane
(Wang et al. 2002b, H\"oflich et al. 2001).

\begin{figure}[ht]
\includegraphics[width=3.2cm,angle=270,clip=]{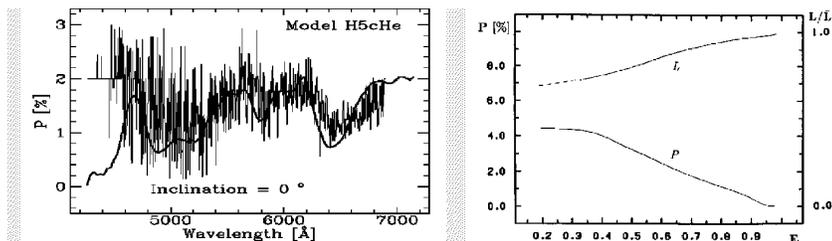}
\vskip -0.4cm
\caption {
Polarization spectrum for SN1993J (Trammel et al. 1995)  in comparison to a model spectrum using
an oblate ellipsoid with r an axis ratio of 1/2.
(left plot). On the right,
the dependence of the continuum polarization (right) and directional
dependence of the luminosity is shown 
 as a function
axis ratios for oblate  ellipsoids  seen from the equator
(from H\"oflich, 1991 \& H\"oflich et al. 1995).
}
\vskip -0.5cm
\label{pol2}
\end{figure}
\begin{figure}[ht]
\includegraphics[width=8.0cm,angle=270,clip=]{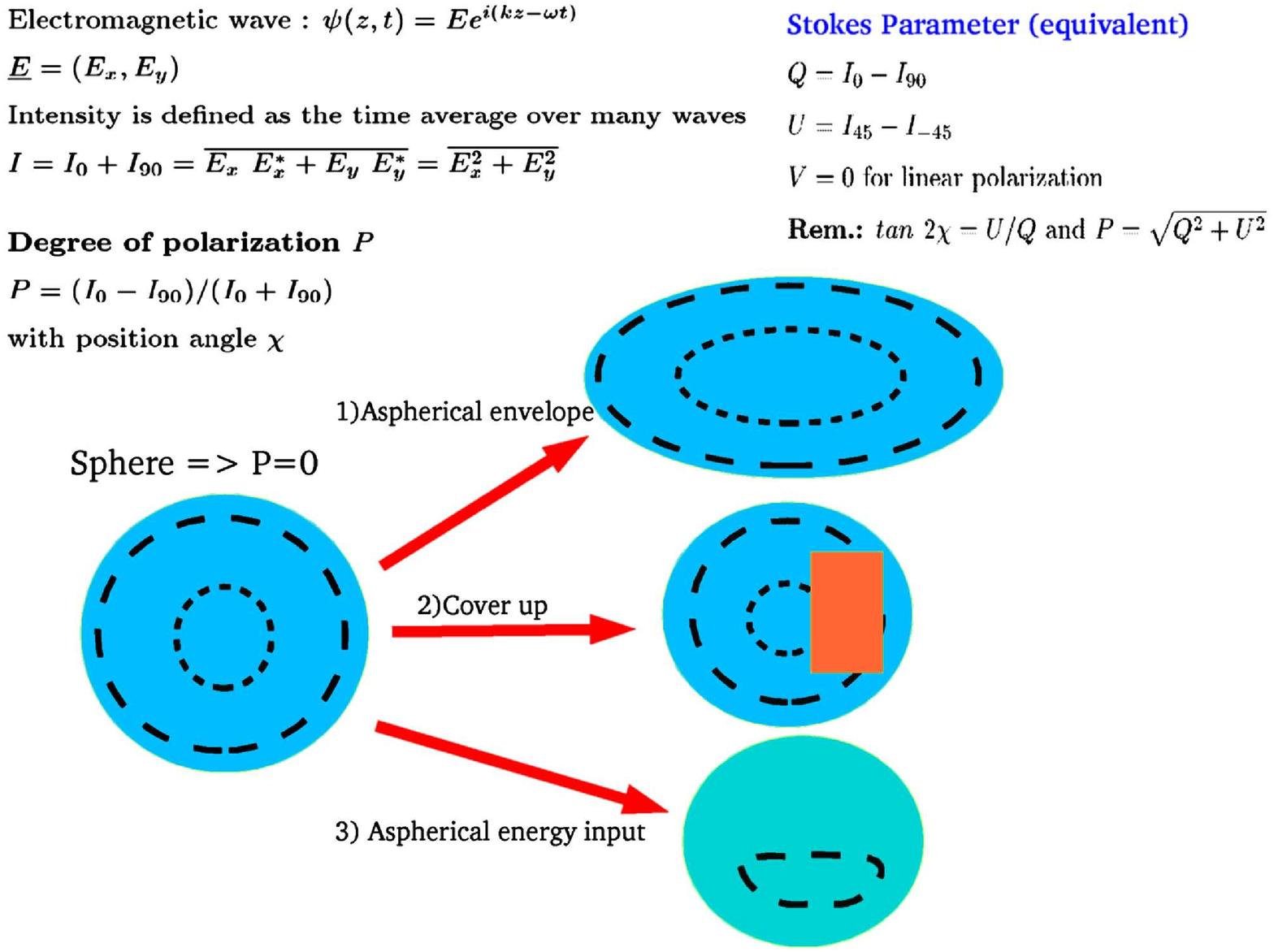}
\vskip -0.5cm
\caption {
 Definition of the polarization and schematic diagram for its production.
 The dotted lines give the
main orientation of the electrical vectors.
 For an unresolved sphere, the components
cancel  out.  Three main mechanisms can be distingished. $\bar P$ can be caused
 by 1) an aspherical envelope, 2)  shading  parts of the disk, or by
3) an aspherical excitation/ionization.
}
\vskip -0.7cm
\label{cases}
\end{figure}

\section{Jet-Induced Supernovae and  Observations}
\vskip -0.2cm
 In supernovae, electron scattering is the main mechanism to  polarize the light.
It can be caused by asymmetries in the density, abundances or excitation structure of an envelope.
In general, the supernovae ejecta cannot be spatially resolved. Although the light from different parts
of a spherical disk is polarized, the resulting polarization $\bar P$ is zero for the integrated light (Fig. 2).
 To produce $\bar P$, three basic configurations must be considered. The envelope is aspherical, parts of    
 the disk are  shaded, and the envelope may be illuminated by an off-center light source.
 In case 2, the shading may be either by a broad-band  absorber such as dust or a specific line opacity.
In the latter case, this would produce a change of $\bar P$ in a narrow line range(Figs. 4). In reality, a combination
of all cases are realized (see below). Note that quantitative analyses of SNe (Fig. 1) need to take into
account that the continua and lines are formed in the same layers.

We have numerically studied the explosion of core collapse
supernovae caused by supersonic jets generated in the center of the star to ask for the jet-properties needed
to reproduce observations.
 The initial stellar structures are based on stellar evolution calculations by Straniero et al. (1997).
 The explosion and jet propagation are
calculated by a full  3-D code within a cubic domain.
The Euler equations are  integrated using an explicit, second-order accurate, Godunov type (PPM, Collela 
\& Woodward 1984),
adaptive-mesh-refinement, massively parallel, Fully-Threaded Tree (FTT)
program, ALLA  (Khokhlov 1998).  
 The subsequent evolution, LCs and spectra are calculated by using modules of our
 hydrodynamical radiation transport code for
spherical and full 3-D (HYDRA) (H\"oflich 2002). This code includes modules for
 hydrodynamics using  PPM (without mesh refinement), detailed networks for nuclear processes and
for atomic, non-LTE level populations, and radiation transport (with mesh-refinement).
 Its components have been used
to carry out many of the previous calculations to analyse  light curves, flux and polarization spectra of
thermonuclear and core collapse supernovae ({H\"oflich 1988, ..., 2002).

\noindent
{\bf General results:} We  simulated  the process of the jet propagation through the star,
the redistribution of elements, and radiation transport effects.
 Qualitatively, the jet-induced picture allows to reproduce the 
polarization observed in core collapse supernovae.
 Both asymmetric ionization and density/chemical distributions are 
 crucial for the production of $P$. Even within the picture of jet-induced explosion, the latter effect alone
cannot (!) account for the large $P$ produced in the intermediate H-rich layers
of core-collapse SN with a massive H-rich envelopes (e.g. SN1987A, SN1999em).

 A  strong explosion and a high efficiency for the conversion of the
jet energy requires low jet velocities or a low, initial collimation
of the jet.   With increasing extension of the envelope, the 
conversion factor increases. Typically, we would expect higher
kinetic energies in SNe~II compared to SNe~Ib/c if a significant amount
of explosion energy is carried away by  jets. 
Within the framework of jet-induced SN, the lack of this evidence
suggests low  jet-velocities.
 The He, C, O and Si rich layers  of the 
progenitor show   characteristic, butterfly-shape structures, and jets bring heavy elements (e.g. 
$^{56}Ni$ into the outer layers.
 Due to the high entropies
of the jet material close to the center, this may be a possible site
for r-process elements.  Moreover, aspherical explosion models show a significantly increased fall-back of material  on the
   central object, e.g. a neutron star, on time scales of minutes to hours  which may trigger the delayed formation  of a black hole.
 Fallback and the low velocity material may alter
the  escape probability for $\gamma $-rays produced by radioactive 
decay of $^{56}Ni$ which is critical for mass estimates of $^{56}Ni$ which are based on late time
observations (e.g.
SN98bw).  For details, see Khokhlov et al. (1999) \& H\"oflich et al. (2001).
\begin{figure}[ht]
\vskip -0.7cm
\includegraphics[width=6.8cm,angle=270,clip=]{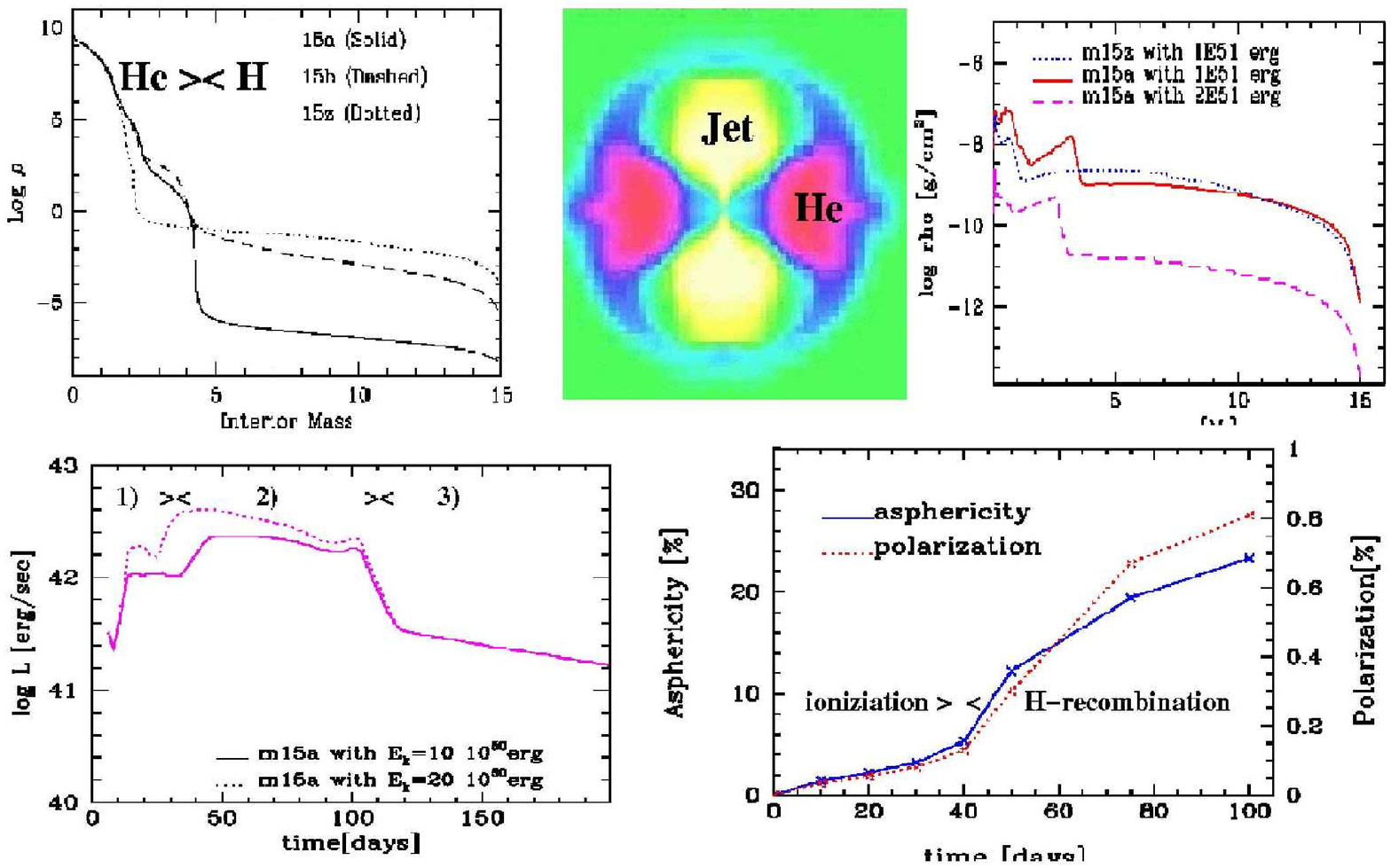}
\vskip -0.2cm
\caption{ Polarization produced by an aspherical, chemical distribution
for an extreme SN~IIp model such as SN1999em (see text).
}
\vskip -0.4cm
\label{93j}
\end{figure}
\begin{figure}[ht]
\includegraphics[width=7.3cm,angle=270,clip=]{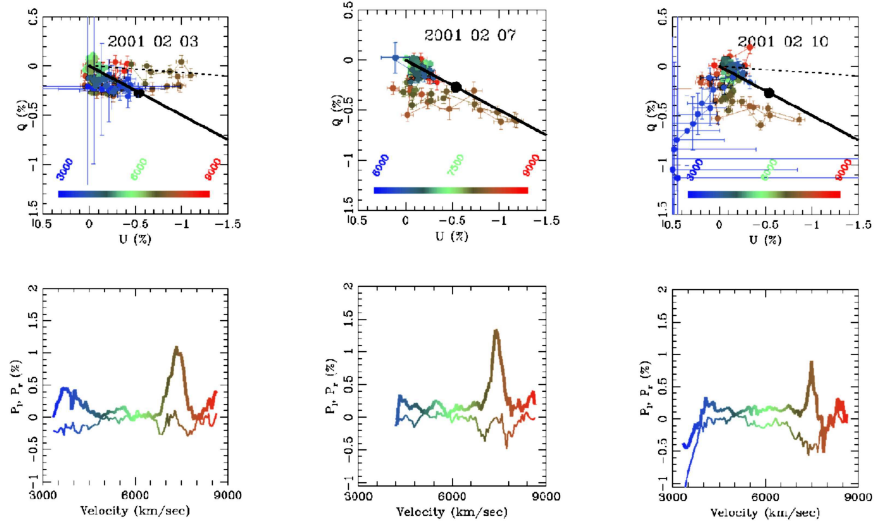}
\vskip -0.3cm
\caption {
Spectropolarimetry of SN2002ap at about -6d, -2d and +1 relative to maximum
light in V. An intrinsic polarization component (shown as a solid dot) is subtracted from the
observed Stokes Parameters so that the data points present the intrinsic
polarization of the SN. The dashed line illustrates the dominant axis
 of the polarization. There has been a distinct shift and, perhaps, a small rotation of the axis in the
Q-U plane during the early epochs, and the spectra are dominated by FeII, NaI D and OI.
 In the spectrum
at about  1d after maximum, the original, dominant axis has shifted further in the Q-U plane, and a new axis
starts to appear at an angle of about $110^o$ defined by Ca II H \&K and Ca II IR triplet. 
 The wavelength color code is presented at the bottom of the panel.
}
\vskip -0.3cm
\label{pol2}
\end{figure}
 
\noindent
{\bf SN1987A and SN1999em:} 
 In our models for these  SNe~II, the jet material stalls within the expanding envelope
corresponding to a velocity of $\approx 4500 km/sec$ during
the phase of homologous expansion.
 In SN1987A, a bump in spectral lines of various elements has been
interpreted by material excited by a clump of radioactive $^{56}Ni$
(Lucy 1988). 
Within our framework, this bump may be a measure of the
region where the jet stalled.  
This could also explain the early appearance of X-rays
 in SN1987A which requires strong mixing of radioactive material
into the hydrogen-rich layers,  and the overall  distribution of elements
in the resolved HST images of the inner debris of SN~1987A.

 For both SN1987A and SN1999em, aspherical excitation by hard radiation  is found to be crucial
to explain the size and presence of the polarization observed early on (Fig. 4).
For the extreme SNIIp 1999em (Fig. 3), our model is based on a star with 15 solar masses and an explosion energy of  2E51erg.
The initial density profile is given for a star at the final stage of stellar evolution
for metalicities Z of 0.02, 0.001 and 0 (models 15a, 15b, 15z, upper left panel)
 For the explosion, we use model 15a.
 In the upper, middle panel, the chemical distribution of He
  is given at 250 sec  for the He-rich layers after the jet material has stalled.
The colors white, yellow, green, blue and red correspond to He mass fractions of
0., 0.18, 0.36, 0.72, and  1., respectively.
 The subsequent explosion has been followed in 1-D up to the phase of homologous
expansion. In the upper, right panel, the density distribution is given at about 5 days
after the explosion.  The steep gradients in the density in the upper right and left panels
are located at the interface between the He-core and the H-mantel.
 In the lower, left panel,
the resulting bolometric LCs are given for explosion energies of
 2E51erg (dotted line) and 1E51erg, respectively.
 Based on full 3-D calculations for the radiation \& $\gamma $-ray transport,
we have calculated the location of the recombination front (in NLTE) as a function of
time. The resulting shape of the photosphere is always prolate.
 The corresponding axis ratio and the  polarization seen from the equator are shown 
 (lower, right panel).
Note the strong increase of the asphericity after the onset of the recombination phase between
day 30 to 40 (H\"oflich et al. 2001). For the polarization in a massive, H-rich envelope,
$P$  is directly liked to the recombination process and asymmetric excitation.

\noindent
{\bf SN2002ap:}
SN~2002ap has attracted much attention because early
spectra showed a lack of hydrogen and helium characteristic of
SN~Ic and broad velocity components (Kinugasa et al. 2002, Meikle et al. 2002, Gal-Yam et al. 2002),
 which have been 
taken as one characteristic of ``hypernovae." The nature,
existence of, and importace of ``hypernovae" remains to be clarified,
and the study of SN~2002ap presents an important opportunity
to shed light on the
general category of ``hypernovae." and their relation to typical SNe~Ic such as SN1994I.

 High-quality spectropolarimetry (range 417-860 nm; spectral resolution 1.27
nm) of SN 2002ap was  obtained with the ESO Very
Large Telescope Melipal (+ FORS1) at 3 epochs that correspond to
-6, -2, and +1 days for a V maximum of 9 Feb 2002.  
The polarization spectra show three distinct
broad ($\sim$ 100 nm) features at $\sim$ 400, 550, and 750 nm that
evolve in shape, amplitude and orientation in the Q-U plane.  
The continuum polarization grows from nearly zero to $\sim$ 0.2 percent.  
The 750 nm feature is polarized at a level $\geq$ 1 percent.
We identify the 550 and 750 nm features as Na I D  
and OI $\lambda$ 777.4 moving at about 20,000 $km/sec$.  
The blue feature may be Fe II.
We interpret the polarization evolution in terms of the impact of 
a bipolar flow from the core that is stopped within the outer envelope of 
a carbon/oxygen core and, consequently,  the Ca features show 
up only at about maximum light. The interpretation of a stalled jet is also supported by IR-spectra taken
by C. Gerardy and M. Meikle (2002, private communications) which show strong CI lines (940.5 and 1070. nm) at expansion velocities
 of $\approx ~ 15,000$ to $ 25,000 km/sec$
 but not, as in SNe~Ia, the strong   1600 to 1900 $nm$ feature due to Fe/Co/Ni.
 Although the symmetry axis remains fixed,
the photosphere retreats by different amounts in different 
directions due to the asymmetric velocity flow and density distribution 
geometrical blocking effects leading to a continuous shift with time of the
main axis of polarization. At about maximum light, the appearance of an additional 
axis in a Q-U plane due to Ca and  processed material indicates a second axis of symmetry.
 Qualitatively and within the picture of jet-induced supernovae this may be explained by
bipolar-jets which are not perfectly aligned and, thus, produce  a kick of the central region
(e.g. neutron star and processed material). Detailed radiation-hydro calculations are under way.
We conclude that the features that characterize SN~2002ap, specifically
its high velocity, can be accounted for in an asymmetric model with
a larger ejecta mass than SN~1994I such that the photosphere remains
longer in higher velocity material.  The characteristics of 
``hypernovae" may be the result of orientation effects in a mildly 
inhomogeneous set of progenitors, rather than requiring an excessive 
total energy or luminosity.  In the analysis of asymmetric events
with spherically symmetric models, it is probably advisable to
refer to ``isotropic equivalent" energy, luminosity, ejected mass,
and nickel mass. This aspect may also be relevant for the interpretation of the
hypernova SN1998bw.

\acknowledgements
       This work was supported in part by NASA Grant NAG5-7937 to PAH.

\end{document}